\documentclass[11pt]{article}
\usepackage{amsmath,amssymb,color,graphics,epsfig,cite}

\usepackage{amsfonts}

\newcommand{\be}{\begin{equation}}
\newcommand{\ee}{\end{equation}}
\newcommand{\bea}{\setlength\arraycolsep{2pt} \begin{eqnarray}}
\newcommand{\eea}{\end{eqnarray}}
\newcommand{\nn}{\nonumber}

\def\0{{\sst{(0)}}}
\def\1{{\sst{(1)}}}
\def\2{{\sst{(2)}}}
\def\3{{\sst{(3)}}}
\def\4{{\sst{(4)}}}
\def\5{{\sst{(5)}}}
\def\6{{\sst{(6)}}}
\def\7{{\sst{(7)}}}
\def\8{{\sst{(8)}}}
\def\sst#1{{\scriptscriptstyle #1}}
\def\oneone{\rlap 1\mkern4mu{\rm l}}

\begin{document}

\begin{center}
{\Large {\bf Melvin-type Electromagnetized Kerr Black Hole in Kaluza-Klein and Einstein-Maxwell-dilaton-axion Theories}}

\vspace{20pt}

Liang Ma, Zhi-Chao Li and H. L\"u

\vspace{10pt}

{\it Center for Joint Quantum Studies, Department of Physics,\\
School of Science, Tianjin University, Tianjin 300350, China }

\vspace{40pt}

\underline{ABSTRACT}
\end{center}

A naive application of solution-generating techniques to construct Melvin-type rotating black holes in the Kaluza-Klein (KK) theory in the presence of both electric and magnetic external fields generically introduces an undesirable Taub-swirling twist term. We investigate its origin and modify the electromagnetization procedure, thereby constructing an exact solution of the Melvin-type electromagnetized Kerr black hole in the KK theory without the swirling term. We also obtain the analogous solution in the Einstein-Maxwell-dilaton-axion (EMDA) theory. The modified solution-generating techniques allow one to introduce both the dyonic charges and external electromagnetic fields independently into Ricci-flat seed metrics in both theories, without generating either Taub-NUT or Taub-swirling twists.

\vfill {\footnotesize maliang0@tju.edu.cn\ \ \ lizc@tju.edu.cn\ \ \ mrhonglu@gmail.com}


\thispagestyle{empty}
\pagebreak

\tableofcontents

\newpage

\section{Introduction}

Kaluza-Klein (KK) theory \cite{Kaluza:1921tu,Klein:1926tv}, developed over a century ago, plays an important role in both classical gravity and string theory, which is inherently higher dimensional. The theory is closely related to Einstein-Maxwell theory, as the dyonic Gibbons-Wiltshire black hole \cite{Gibbons:1985ac} naturally reduces to the corresponding Reissner-Nordstr\"om black hole when its electric and magnetic charges are equal in magnitude. Remarkably, the equations governing the Gibbons-Wiltshire black hole can be recast as a one-dimensional Toda equation of $SL(3,\mathbb R)$.
This provides an early example of the connection between black hole and $p$-brane solutions and Liouville and Toda equations, which has been extensively explored in subsequent studies \cite{Lu:1996hh,Lu:1996jr,Lu:1996jk,Lu:1997rd,Goldstein:2005hq,Lu:2013uia,Ivashchuk:2002ge, Ivashchuk:2013jja,Zadora:2017qeo,Lu:2026lkv}. More recently, it has been shown that the thermodynamics of these black holes can be derived without explicitly constructing the solutions after all \cite{Yang:2025rud,Lu:2025eub,Han:2026bim, Wu:2026udi,Huang:2026ivj}.

The simple relation between five-dimensional vacuum gravity and the four-dimensional KK theory provides a natural five-to-four route of generating new solutions. The electrically-charged KK black hole is simply a boosted four-dimensional Schwarzschild black hole along the internal direction \cite{Dobiasch:1981vh,Chodos:1980df}. Its magnetically-charged counterpart can be oxidized to five dimensions to become the KK monopole \cite{Sorkin:1983ns,Gross:1983hb}. However, boosting the KK monopole does not yield the expected Gibbons-Wiltshire black hole, but instead produces a dyonic solution with a Taub-NUT twist \cite{Taub:1950ez,Newman:1963yy}, proportional to the product of electric and magnetic charges. Without an appropriate solution-generating technique, constructing the rotating generalization directly from the equations of motion would be extremely difficult. This difficulty was overcome independently by Rasheed and Larsen. Their construction exploits the four-to-three dimensional reduction under which the KK theory gives rise to a three-dimensional sigma model with a global $SL(3, \mathbb R)$ symmetry. Selecting a suitable $SL(3,\mathbb R)$ transformation on a suitable Ricci-flat seed yields not only the Gibbons-Wiltshire solution, but also its rotating generalization, known as the Rasheed-Larsen black hole \cite{Rasheed:1995zv,Larsen:1999pp}.

In this paper, we consider the Melvin-type magnetized solutions in the KK theory \cite{Dowker:1994up,Dowker:1995gb}, with an aim of constructing black holes immersed in both electric and magnetic external fields, characterized with parameters $E$ and $B$ respectively. The corresponding electromagnetized Minkowski vacuum in KK theory was already known \cite{Ivashchuk:2002jr, Bolokhov:2017rnf}. The electromagnetized Schwarzschild black hole in KK theory was recently constructed \cite{Ma:2026rro}; but constructing its rotating generalization directly from the equations of motion remains challenging.\footnote{It is our understanding that Hai Huang is developing a new technique for constructing such a solution directly from the equations of motion in Einstein-Maxwell-dilaton theories.} A naive application of the solution-generating technique via the five-to-four route introduces a Taub-swirling \cite{Taub:1950ez}, the electric-magnetic counterpart of the Taub-NUT geometry, when both $E$ and $B$ are turned on. We therefore explore the four-to-three route and identify an appropriate $SL(3,\mathbb R)$ transformation that generates the Schwarzschild and Kerr black holes immersed in the electromagnetic fields without swirling. This also helps us to modify the five-to-four route to generate the same solutions without swirling. Moreover, we show that the Rasheed-Larsen solution can also be obtained directly through this modified five-to-four route, without going down to three dimensions. Beyond KK theory, we also construct an exact solution describing the Melvin-type of Kerr black hole immersed in external electromagnetic fields in Einstein-Maxwell-dilaton-axion (EMDA) theory.

The paper is organized as follows. In Section \ref{sec:emvacuum}, we focus on electromagnetized vacuum using solution generating technique, either via the five-to-four route or via the four-to-three route. We examine the origin of the swirling twist and how it can be eliminated in both approaches, and construct a static electromagnetized vacuum free of the swirling term. We obtain the static electromagnetized vacuum without swirling. In Section \ref{sec:Melvin-Kerr}, we apply the electromagnetization procedure to the Kerr black hole and obtain the corresponding solution without swirling. In Section \ref{sec:emda}, we extend our analysis to EMDA theory and obtain the corresponding electromagnetized Kerr black hole. We conclude in Section \ref{sec:conclusions}. In Appendix \ref{sec:kkdyon}, we derive the Gibbons-Wiltshire black hole using a modified five-to-four solution-generating technique, without going down to three dimensions. In Appendix \ref{sec:nutswirling}, we employ suitably chosen $SL(3,\mathbb R)$ transformations to generate both the Ricci-flat Taub-NUT and Taub-swirling metrics.

\section{Static electromagnetized vacuum in KK teory}
\label{sec:emvacuum}

The KK theory is a particular case of Einstein-Maxwell-dilaton (EMD) gravity, arising from the KK circle reduction of five-dimensional pure gravity, with the complete massless sector retained. The reduction ansatz from five to four dimensions is
\be
d\hat{s}^2=e^{-\frac{\phi}{\sqrt{3}}}ds_4^2+e^{\frac{2\phi}{\sqrt{3}}}(dy+A_{\1})^2\,.\label{KK reduction 5 to 4}
\ee
The resulting four-dimensional theory contains the metric, the KK vector $A_\1$ and a dilaton $\phi$. The Lagrangian is
\be
\mathcal{L}=\sqrt{-g}\Big( R-\frac{1}{2}(\partial\phi)^2-\frac{1}{4}e^{\sqrt{3}\phi}F_\2^2\Big)\,,\qquad F_{\2}=dA_{\1}\,.\label{KK EMD}
\ee
Equivalently, the theory may also be reformulated in the language of differential forms:
\be
\mathcal{L}=R\, {*\oneone}-\frac{1}{2}{*d}\phi\wedge d\phi-\frac{1}{2}e^{\sqrt{3}\phi}{*F}_{\2}\wedge F_{\2}\,,
\ee
where $*$ denotes the Hodge dual in four dimensions. When the matter fields $\phi=0$ and $A_{\1}=0$, the KK theory reduces to pure Einstein gravity. Furthermore, the ansatz \eqref{KK reduction 5 to 4} implies that any four-dimensional Ricci-flat geometry can be uplifted to a five-dimensional Ricci-flat metric, by taking a direct product with an additional circle (or $\mathbb R$) direction, corresponding to $dy^2$. Consequently, four-dimensional Ricci-flat metrics are often commonly used as seed solutions for generating new solutions in the KK theory.

\subsection{Swirling Melvin universe from 5D }

We now consider a specific solution-generating technique in the KK theory \cite{Dowker:1994up, Dowker:1995gb}, based on the KK reduction ansatz \eqref{KK reduction 5 to 4}. Starting from a four-dimensional Ricci-flat metric $ds_4^2$, we first uplift it trivially to construct a five-dimensional Ricci-flat geometry $ds_5^2=ds_4^2+dy^2$. We then perform the coordinate transformation $\varphi\rightarrow\varphi'=\varphi+By$, where $\varphi$ is a spacelike Killing coordinate of $ds_4^2$. Finally, reducing along the $y$ direction yields a Melvin-type magnetized solution. We denote to
this entire magnetization procedure by $M(B)$, which consists of three steps: uplifting, the coordinate transformation and a final dimensional reduction.

Building upon the magnetized solution, we can introduce the concept of electricization. We define the electromagnetic duality transformation $\mathcal{D}$ and its inverse $\mathcal{D}^{-1}$ as follows
\bea
\mathcal{D}:&&\qquad \phi\rightarrow -\phi\,,\qquad F_{\2}\rightarrow e^{\sqrt{3}\phi}{*F}_{\2}\,,\cr
\mathcal{D}^{-1}:&&\qquad \phi\rightarrow -\phi\,,\qquad F_{\2}\rightarrow -e^{\sqrt{3}\phi}{*F}_{\2}\,.
\label{ele mag dual}
\eea
The equations of motion are invariant under both duality transformations. The discrete electromagnetic duality symmetry makes the construction of dyonic black holes or black holes immersed in both electric and magnetic external fields a more challenging task.

The electricization procedure that causes a Ricci-flat black hole immersed in the external electric field is simply magnetization followed by performing electromagnetic duality or its inverse, namely, $\mathcal{D}\circ M(E)$ or $\mathcal{D}^{-1}\circ M(E)$. The full electromagnetization process that both electricizes and magnetizes a Ricci-flat metric is then given by $N(E)\circ M(B)$, where
\be
N(E)=\mathcal{D}\circ M(E)\circ \mathcal{D}^{-1}\,.
\ee
In other words, starting from the magnetized solution with parameter $B$, we then apply the inverse electromagnetic duality transformation $\mathcal{D}^{-1}$, which maps the external magnetic field to its electric counterpart with the same parameter $B$. We then apply the magnetization procedure again, but with parameter $E$, followed by the electromagnetic duality transformation $\mathcal{D}$. This yields a solutions immersed in both external electric and magnetic fields, characterized by the parameters $E$ and $B$ respectively.

We now apply the above procedures to the four-dimensional Minkowski spacetime in cylindrical coordinates
\be
ds_4^2=-dt^2+d\rho^2+dz^2+\rho^2d\varphi^2\,.\label{Mink}
\ee
The magnetized solution under $M(B)$ is given by
\bea
ds_{4,\1}^2&=&\sqrt{H_1}(-dt^2+d\rho^2+dz^2)+\frac{\rho^2}{\sqrt{H_1}}d\varphi^2\,,\qquad A_{\1}^{\1}=\frac{B\rho^2}{H_1}d\varphi\,,\cr
\phi^{\1}&=&\frac{\sqrt{3}}{2}\log H_1\,,\qquad H_1=1+B^2\rho^2\,.
\eea
The next electricization procedure $N(E)$ consists of three steps. First, the inverse electromagnetic duality transformation $\mathcal{D}^{-1}$ leaves the metric invariant while transforming the Maxwell field and the dilaton as follows
\be
ds_{4,\2}^2=\sqrt{H_1}(-dt^2+d\rho^2+dz^2)+\frac{\rho^2}{\sqrt{H_1}}d\varphi^2\,,\quad A_{\1}^{\2}=-2Bzdt\,,\quad
\phi^{\2}=-\frac{\sqrt{3}}{2}\log H_1\,.
\ee
We then perform magnetization $M(E)$ with parameter $E$, yielding the following solution
\bea
ds^2_{4,\3}&=&\sqrt{H_1H_2}(-dt^2+d\rho^2+dz^2)+\frac{\rho^2}{\sqrt{H_1H_2}}(d\varphi+2BEzdt)^2\,,\cr
A_{\1}^{\3}&=&-2Bzdt+\frac{E\rho^2H_1}{H_2}(d\varphi+2BEzdt)\,,\qquad \phi=\frac{\sqrt{3}}{2}\log\frac{H_2}{H_1}\,,\cr
H_1&=&1 + B^2 \rho^2\,,\qquad H_2=1+E^2\rho^2+B^2E^2\rho^4\,.\label{H1H2}
\eea
In this solution, the electric and magnetic fields are characterized by $B$ and $E$ respectively. We therefore finally apply the electromagnetic duality and the final resulting solution is
\bea
ds^2&=&\sqrt{H_1H_2}(-dt^2+d\rho^2+dz^2)+\frac{\rho^2}{\sqrt{H_1H_2}}(d\varphi+2BEzdt)^2\,,\cr
A_{\1}&=&2Ezdt+\frac{B\rho^2}{H_1}(d\varphi+2BEzdt)\,,\qquad \phi=\frac{\sqrt{3}}{2}\log\frac{H_1}{H_2}\,,\label{Mink swirling}
\eea
where $(H_1,H_2)$ are given in \eqref{H1H2}. It is evident from the above expressions that the solution reduces to a Melvin-type magnetic vacuum when $E=0$, and to its electric counterpart when $B=0$. However, when both $B$ and $E$ are nonzero, the metric develops a nontrivial cross term $dt\,d\varphi$, associated with a structure known as the Taub swirling. Although the resulting configuration is indeed an exact solution of the KK theory, this swirling structure renders it non-static. On the other hand, the static vacuum with both external electric and magnetic fields does exist in the KK theory, and it was constructed as a special case of more general EMD theories \cite{Bolokhov:2017rnf,Ma:2026rro}. This raises a question: can we develop a solution generating technique to derive the solution without the swirling? We shall resolve this issue in the next subsection.

\subsection{Hidden $SL(3,\mathbb{R})$ symmetry and static electromagnetized vacuum}

The hidden $SL(3,\mathbb{R})$ symmetry of the KK theory \eqref{KK EMD} becomes manifest upon dimensional reduction. The origin of this symmetry can be understood as follows. Einstein gravity reduced on $n$-torus has $GL(n,\mathbb R)$ global symmetry and if the resulting theory is in three dimensions, the symmetry is enhanced to $SL(n+1,\mathbb R)$ \cite{Cremmer:1997ct,Cremmer:1999du}. The $SL(3,\mathbb R)$ symmetry of KK theory underlies the Ernst transformations, which provide a powerful solution-generating technique. Applying an electric Ernst transformation to the Kerr black hole as a seed solution generates a rotating dyonic black hole, known as the Rasheed-Larsen solution \cite{Rasheed:1995zv,Larsen:1999pp}. Since our aim is to introduce external electric and magnetic fields rather than electric and magnetic charges, we instead employ a magnetic Ernst transformation. Specifically, we adopt the following standard dimensional reduction ansatz\footnote{To generate dyonic charges, the dimensional reduction must instead be performed along the time direction.}
\be
ds_4^2=e^{\Phi}(dz+\bar{\mathcal{A}}_{\1})^2+e^{-\Phi}d\bar{s}_3^2\,,\qquad A_{\1}=\bar{A}_{\1}+\chi(dz+\bar{\mathcal{A}}_{\1})\,.\label{magnetic Ernst}
\ee
Introducing the following notation
\be
F_{\2}=dA_{\1}=\bar{F}_{\2}+d\chi\wedge(dz+\bar{\mathcal{A}}_{\1})\,,\qquad
\bar{F}_{\2}=d\bar{A}_{\1}+\chi\bar{\mathcal{F}}_{\2}\,,\qquad \bar{\mathcal{F}}_{\2}=d\bar{\mathcal{A}}_{\1}\,,
\ee
we obtain the three-dimensional Lagrangian
\bea
\bar{\mathcal{L}}&=&\bar{R}\,{\bar{*}\oneone}-\frac{1}{2}\,\bar{*}d\Phi\wedge d\Phi-\frac{1}{2}\,\bar{*}d\phi\wedge d\phi-\frac{1}{2}e^{\sqrt{3}\phi-\Phi}\,\bar{*}d\chi\wedge d\chi\cr
&&-\frac{1}{2}e^{2\Phi}\,\bar{*}\bar{\mathcal{F}}_{\2}\wedge \bar{\mathcal{F}}_{\2}-\frac{1}{2}e^{\sqrt{3}\phi+\Phi}\,\bar{*}\bar{F}_{\2}\wedge \bar{F}_{\2}\,,\label{3D action}
\eea
where $\bar{*}$ is the Hodge dual in three dimensions. We recast the three-dimensional action in first-order form by introducing two additional scalar fields, \(\{\sigma,\psi\}\), as Lagrange multipliers to impose the Bianchi identities
\be
d\bar{\mathcal{F}}_{\2}=0\,,\qquad d\bar{F}_{\2}-d\chi\wedge\bar{\mathcal{F}}_{\2}=0\,.\label{Bianchi}
\ee
The resulting first-order Lagrangian is
\bea
\bar{\mathcal{L}}&=&\bar{R}\,\bar{*}\oneone-\frac{1}{2}\,\bar{*}d\Phi\wedge d\Phi-\frac{1}{2}\,\bar{*}d\phi\wedge d\phi-\frac{1}{2}e^{\sqrt{3}\phi-\Phi}\,\bar{*}d\chi\wedge d\chi\cr
&&-\frac{1}{2}e^{2\Phi}\,\bar{*}\bar{\mathcal{F}}_{\2}\wedge \bar{\mathcal{F}}_{\2}-\frac{1}{2}e^{\sqrt{3}\phi+\Phi}\,\bar{*}\bar{F}_{\2}\wedge \bar{F}_{\2}
+\sigma (d\bar{F}_{\2}-d\chi\wedge \bar{\mathcal{F}}_{\2})+\psi d\bar{\mathcal{F}}_{\2}\,.\label{3D action 1}
\eea
Varying the Lagrangian with respect to the two two-form field strengths, we obtain the corresponding equations of motion
\be
e^{\sqrt{3}\phi+\Phi}\,\bar{*}\bar{F}_{\2}+d\sigma=0\,,\qquad e^{2\Phi}\,\bar{*}\bar{\mathcal{F}}_{\2}+d\psi+\sigma d\chi=0\,.
\ee
Solving for the two two-forms and substituting them into \eqref{3D action 1}, we arrive at a three-dimensional theory described entirely by five scalar fields, namely
\bea
\bar{\mathcal{L}}&=&\bar{R}\,\bar{*}\oneone-\frac{1}{2}\,\bar{*} d\Phi\wedge d\Phi-\frac{1}{2}\,\bar{*} d\phi\wedge d\phi-\frac{1}{2}e^{\sqrt{3}\phi-\Phi}\,\bar{*} d\chi\wedge d\chi\cr
&&-\frac{1}{2}e^{-\sqrt{3}\phi-\Phi}\,\bar{*} d\sigma\wedge d\sigma-\frac{1}{2}e^{-2\Phi}\,\bar{*} (d\psi+\sigma d\chi)\wedge(d\psi+\sigma d\chi)\,.\label{3D action scalar}
\eea
The five scalar fields parametrize the symmetric space \(SL(3,\mathbb{R})/SO(3)\). This structure becomes manifest upon introducing a symmetric matrix \(S\)
\begin{equation}
\begingroup
\renewcommand{\arraystretch}{1.8}
\setlength{\arraycolsep}{10pt}
S=e^{\frac{\phi}{\sqrt{3}}}
\begin{pmatrix}
e^{-\Phi}
& e^{-\Phi}\chi
& -e^{-\Phi}\psi
\\
e^{-\Phi}\chi
& e^{-\sqrt{3}\phi}+e^{-\Phi}\chi^2
& e^{-\sqrt{3}\phi}\sigma-e^{-\Phi}\chi\psi
\\
-e^{-\Phi}\psi
& e^{-\sqrt{3}\phi}\sigma-e^{-\Phi}\chi\psi
& e^{\Phi}+e^{-\sqrt{3}\phi}\sigma^2+e^{-\Phi}\psi^2
\end{pmatrix},
\endgroup
\end{equation}
which can be further factorized as
\begin{equation}
\begingroup
\renewcommand{\arraystretch}{1.5}
\setlength{\arraycolsep}{8pt}
S=
\begin{pmatrix}
1 & 0 & 0\\
\chi & 1 & 0\\
-\psi & \sigma & 1
\end{pmatrix}
\begin{pmatrix}
e^{-\Phi+\frac{\phi}{\sqrt{3}}} & 0 & 0\\
0 & e^{-\frac{2\phi}{\sqrt{3}}} & 0\\
0 & 0 & e^{\Phi+\frac{\phi}{\sqrt{3}}}
\end{pmatrix}
\begin{pmatrix}
1 & \chi & -\psi\\
0 & 1 & \sigma\\
0 & 0 & 1
\end{pmatrix}.
\endgroup\label{KK EMD S}
\end{equation}
In terms of the matrix \(S\), the Lagrangian \eqref{3D action scalar} can be recast as
\be
\bar{\mathcal{L}}=\bar{R}\, {\bar *}\oneone-\frac{1}{4}\mathrm{tr}[(S^{-1}\,{\bar *}dS )\wedge (S^{-1} dS)]\,.\label{eincoset}
\ee
The scalar sector of the action is invariant under global \(SL(3,\mathbb{R})\) transformations $S\rightarrow S'=\Gamma S\Gamma^T$, where $\Gamma$ is a \(3\times 3\) real constant matrix with \(\det\Gamma=1\). Note that $S^T=S$ and $\det S=1$, and the $SL(3,\mathbb R)$ transformation preserves these properties. It follows that solutions to $S'=\Gamma S\Gamma^T$ must exist for any $SL(3,\mathbb R)$ matrix $\Gamma$. For our electromagnetization procedure, we find that the appropriate transformation is generated by the matrix:
\be
\Gamma(B,E)
=
\renewcommand{\arraystretch}{1.5}
\setlength{\arraycolsep}{12pt}
\begin{pmatrix}
1 & B & \dfrac{BE}{2} \\
0 & 1 & E \\
0 & 0 & 1
\end{pmatrix}\,.\label{SL3R BE}
\ee
The necessity of including the upper-right entry in the upper-triangular matrix is analogous to the role played by the corresponding $SU(2,1)$ transformation in Einstein-Maxwell theory \cite{Gibbons:2013yq}. Applying the above transformations, the four-dimensional Minkowski spacetime is transformed into a static electromagnetized vacuum
\bea
ds^2&=&\sqrt{H_1H_2}(-dt^2+d\rho^2+dz^2)+\frac{\rho^2}{\sqrt{H_1H_2}}d\varphi^2\,,\cr
A_{\1}&=&2Ezdt+\frac{B\rho^2}{H_1}\Big(1+\frac{1}{2}E^2\rho^2\Big)d\varphi\,,\qquad \phi=\frac{\sqrt{3}}{2}\log\frac{H_1}{H_2}\,,\cr
H_1&=&1+B^2\rho^2+\frac{1}{4}B^2E^2\rho^4\,,\qquad H_2=1+E^2\rho^2+\frac{1}{4}B^2E^2\rho^4\,.\label{Mink BE}
\eea
This is the electromagnetized vacuum without swirling, and was constructed directly \cite{Bolokhov:2017rnf,Ma:2026rro}. When $E=B$, the dilaton decouples and the solution reduces to the Melvin universe with equal electric and magnetic external fields.

Next, we examine in detail the $SL(3,\mathbb{R})$ transformation matrix $\Gamma$ in \eqref{SL3R BE}, which generates the desired configuration with both external electric and magnetic fields. The magnetization and electricization procedures introduced in the previous subsection are both $SL(3,\mathbb{R})$ transformations, represented by the following matrices, respectively
\be
M(B)
=
\renewcommand{\arraystretch}{1.15}
\setlength{\arraycolsep}{7pt}
\begin{pmatrix}
1 & B & 0 \\
0 & 1 & 0 \\
0 & 0 & 1
\end{pmatrix}\,,\qquad
N(E)
=
\begin{pmatrix}
1 & 0 & 0 \\
0 & 1 & E \\
0 & 0 & 1
\end{pmatrix}\,.\label{MBNE}
\ee
The magnetization transformation matrix $M(B)$ was previously given in \cite{Yazadjiev:2013hxa}.

The procedure described in the previous section, which consists of applying magnetization $M(B)$, followed by electricization $N(E)$ to generate the swirling spacetime \eqref{Mink swirling}, can equivalently be realized by the composite $SL(3,\mathbb{R})$ transformation $N(E)M(B)$ in three dimensions. The corresponding transformation is given by
\be
\Gamma'(B,E)
=N(E)M(B)
=
\renewcommand{\arraystretch}{1.15}
\setlength{\arraycolsep}{7pt}
\begin{pmatrix}
1 & B & 0 \\
0 & 1 & E \\
0 & 0 & 1
\end{pmatrix}\,.\label{twist Gamma}
\ee
We observe that, compared to \eqref{SL3R BE}, the transformation matrix \eqref{twist Gamma} is missing the upper-right entry $BE/2$, which accounts for the failure of the procedure in the previous section fails to generate the desired result. To overcome this deficiency, we introduce an additional matrix $\Gamma''(B,E)$
\bea
\Gamma''(B,E)
=
\renewcommand{\arraystretch}{1.15}
\setlength{\arraycolsep}{7pt}
\begin{pmatrix}
1 & 0 & \frac{BE}{2} \\
0 & 1 & 0 \\
0 & 0 & 1
\end{pmatrix}\,,\qquad  \Gamma(B,E)=\Gamma'(B,E)\Gamma''(B,E)\,,\label{Gamma dd}
\eea
to eliminate the swirling structure.

\subsection{$SL(3,\mathbb{R})$ via five-to-four KK reduction and half-magnetization}
\label{Half-Magnetization}

We observe that, through the appropriate \(SL(3,\mathbb{R})\) symmetry \eqref{SL3R BE} of the three-dimensional theory \eqref{3D action scalar}, a reasonable electromagnetized vacuum without swirling \eqref{Mink BE} can be generated. A natural question then arises: can we realize this from the earlier  five-to-four dimensional perspective \eqref{KK reduction 5 to 4}, without going down to three dimensions? The answer is yes. To this end, we introduce the concept of half-strength magnetization.

Note that the five-to-four route allows us to implement only the magnetization $M(B)$ or electricization $N(E)$ procedures defined in \eqref{MBNE}, whereas the transformation $\Gamma''(B,E)$ in \eqref{Gamma dd} cannot be implemented directly. Nevertheless, we find that the transformation $\Gamma(B,E)$ in \eqref{SL3R BE} can be realized through the following sequence of operations
\be
\Gamma(B,E)=M\Big(\frac{B}{2}\Big)N(E)M\Big(\frac{B}{2}\Big)\,.
\ee
Equivalently, starting from a Ricci-flat seed solution, we first perform a half-strength magnetization, followed by electricization, and finally apply another half-strength magnetization, thereby eliminating the swirling structure. Since electricization itself consists of applying the inverse electromagnetic duality, magnetization, and electromagnetic duality transformations, we shall follow this procedure
\be
M\Big(\frac{B}{2}\Big)\circ\mathcal{D}\circ M(E)\circ\mathcal{D}^{-1}\circ M\Big(\frac{B}{2}\Big)
\ee
to construct the solution \eqref{Mink BE} starting from the four-dimensional Minkowski vacuum.

To implement this modified five-to-four route explicitly, we first perform a half-strength magnetization by applying a coordinate transformation with half the magnetic field parameter, $B/2$, to the five-dimensional uplift of Minkowski spacetime, namely
\be
\varphi\qquad \rightarrow\qquad \varphi'=\varphi+\frac{B}{2}y\,.
\ee
We then reduce the resulting geometry to four dimensions by \eqref{KK reduction 5 to 4}, obtaining the solution
\bea
ds_{4,\1}^2&=&\sqrt{\Lambda}(-dt^2+d\rho^2+dz^2)+\frac{\rho^2}{\sqrt{\Lambda}}d\varphi^2\,,\qquad A_{\1}^{\1}=\frac{B\rho^2}{2\Lambda}d\varphi\,,\cr
\phi^{\1}&=&\frac{\sqrt{3}}{2}\log \Lambda\,,\qquad \Lambda=1+\frac{1}{4}B^2\rho^2\,.
\eea
After this step, we have effectively completed a half-strength magnetization $M(B/2)$. Under the electromagnetic duality transformation $\mathcal{D}^{-1}$ in \eqref{ele mag dual}, we convert the magnetic field into an electric field
\be
ds_{4,\2}^2=\sqrt{\Lambda}(-dt^2+d\rho^2+dz^2)+\frac{\rho^2}{\sqrt{\Lambda}}d\varphi^2\,,\qquad A_{\1}^{\2}=-Bzdt\,,\quad
\phi^{\2}=-\frac{\sqrt{3}}{2}\log \Lambda\,.\label{KK step 1}
\ee
Next, we uplift \eqref{KK step 1} to five dimensions and perform a second coordinate transformation to introduce the parameter $E$
\be
\varphi\qquad \rightarrow\qquad \varphi'=\varphi+Ey\,.
\ee
Upon dimensional reduction, we obtain the following four-dimensional solution
\bea
ds_{4,\3}^2&=&\sqrt{\Lambda H_2}(-dt^2+d\rho^2+dz^2)+\frac{\rho^2}{\sqrt{\Lambda H_2}}(d\varphi+BEzdt)^2\,,\cr
A_{\1}^{\3}&=&-Bzdt+\frac{E\rho^2\Lambda}{H_2}(d\varphi+BEzdt)\,,\qquad \phi^{\3}=\frac{\sqrt{3}}{2}\log \frac{H_2}{\Lambda}\,,\cr
H_2&=&1+E^2\rho^2+\frac{1}{4}B^2E^2\rho^4\,.
\eea
A nontrivial swirling term has already emerged at this stage. We then apply the electromagnetic duality transformation $\mathcal{D}$ in \eqref{ele mag dual} once again, obtaining
\bea
ds_{4,\4}^2&=&\sqrt{\Lambda H_2}(-dt^2+d\rho^2+dz^2)+\frac{\rho^2}{\sqrt{\Lambda H_2}}(d\varphi+BEzdt)^2\,,\cr
A_{\1}^{\4}&=&2Ezdt+\frac{B\rho^2}{2\Lambda}(d\varphi+BEzdt)\,,\qquad \phi^{\4}=\frac{\sqrt{3}}{2}\log \frac{\Lambda}{H_2}\,.
\eea
Finally, we perform another half-strength magnetization $M(B/2)$ on this solution, yielding precisely the correct static electromagnetic vacuum \eqref{Mink BE}. Thus, we establish that the electromagnetic vacuum can be generated through the modified five-to-four route without going down to the three dimensions.

\section{Electromagnetized Kerr black hole in KK theory}
\label{sec:Melvin-Kerr}

Having established the solution-generating procedures for constructing electromagnetized vacuum in the KK theory, with or without swirling, we now apply these procedures to the Kerr metric
\bea
ds^2&=&-\Big(1-\frac{2\mu r}{\Sigma}\Big)dt^2-\frac{4\mu a r(1-x^2)}{\Sigma}dtd\varphi+\Sigma\frac{dr^2}{\Delta}\cr
&&+\Sigma\frac{dx^2}{1-x^2}+\Big(r^2+a^2+\frac{2\mu a^2r(1-x^2)}{\Sigma}\Big)(1-x^2)d\varphi^2\,,\cr
\Delta&=&r^2-2\mu r+a^2\,,\qquad \Sigma=r^2+a^2x^2\,,\qquad x=\cos\theta\,.\label{Kerr}
\eea
The solutions can be constructed through either the four-to-three route or equivalently the modified five-to-four route. The final answer is the same. The electromagnetized Kerr black hole without swirling in the KK theory is given by
\bea
ds^2&=&\sqrt{H_1H_2}\Sigma\Big[
-\frac{\Delta}{\mathbb{P}}dt^2+\frac{dr^2}{\Delta}+\frac{dx^2}{1-x^2}
\Big]+\frac{(1-x^2)\mathbb{P}}{\sqrt{H_1H_2}\Sigma}(d\varphi+\mathbb{T}dt)^2\,,\nn\\
A_{\1}&=&\frac{1}{H_1}\Big[\frac{2}{B}(H_1-1)+E\mathbb{L}-
\frac{B(1-x^2)\mathbb{P}}{\Sigma}\Big](d\varphi
+\mathbb{T}dt)-E\Big[\mathbb{Z}-\frac{BE}{2}\mathbb{M}\Big]dt\,,\cr
\phi&=&\frac{\sqrt{3}}{2}\log\frac{H_1}{H_2}\,,\label{Kerr BE}
\eea
where
\bea
H_1&=&\Big(1-\frac{1}{2}BE\mathbb{L}\Big)^2+B^2\frac{(1-x^2)\mathbb{P}}{\Sigma}
+\frac{1}{4}B^2E^2\Big(\frac{(1-x^2)\mathbb{P}}{\Sigma}\Big)^2\,,\cr
H_2&=&\Big(1+\frac{1}{2}BE\mathbb{L}\Big)^2+E^2\frac{(1-x^2)\mathbb{P}}{\Sigma}
+\frac{1}{4}B^2E^2\Big(\frac{(1-x^2)\mathbb{P}}{\Sigma}\Big)^2\,,\cr
\mathbb{T}&=&-\frac{2\mu a r}{\mathbb{P}}-\frac{1}{4}B^2E^2\mathbb{M}\,,\qquad \mathbb{Z}=-\frac{2x\Delta}{\mathbb{P}}\big[r\Sigma-\mu a^2(1+x^2)\big]\,,\cr
\mathbb{M}&=&\frac{2\mu a}{\mathbb{P}}\Big[
16 \mu ^2 a^2 r+\Delta  \big(a^2 r (1-6 x^2-3 x^4)-r^3 (3+6 x^2-x^4)+2 \mu  a^2 (1+6 x^2+x^4)\big)
\Big]\,,\cr
\mathbb{P}&=&(r^2+a^2)^2-a^2(1-x^2)\Delta\,,\qquad \mathbb{L}=-2\mu ax\Big(3-x^2+\frac{a^2(1-x^2)^2}{\Sigma}\Big)\,.\label{Kerr BE solution}
\eea
In the static limit $a=0$, the solution reduces to the electromagnetized Schwarzschild black hole in KK theory, which was constructed directly from the equations of motion \cite{Ma:2026rro}. Setting further $\mu=0$ yields the static electromagnetized vacuum discussed earlier with $z=r x$ and $\rho=r \sqrt{1-x^2}$. In general, to eliminate the conical singularity at either north or south pole, we need to rescale our azimuthal coordinate $\varphi$ by
\be
\varphi\quad \rightarrow\quad \varphi'=P_0\,\varphi\,,\qquad P_0=1-4\mu^2a^2B^2E^2\,.
\ee
so that the resulting $\varphi$ have a $2\pi$ period.

However, our solution cannot be regarded as a genuine electromagnetized Kerr black hole, in the sense of
\cite{Ovcharenko:2026tos}. It turns out that the solution is not neutral, and dyonic charges emerge via the rotation. Specifically, we find that the electric and magnetic charges are respectively given by
\be
Q_e=\frac{1}{16\pi}\int e^{\sqrt{3}\phi}{*F}_{\2}=\mu aB\,,\qquad Q_m=\frac{1}{16\pi}\int F_{\2}=\mu aE\,.\label{dyoniccharges}
\ee
It can be seen that both charges are not independent, but arising entirely from the rotation itself, analogous to the situation for the Kerr-Melvin black hole in Einstein-Maxwell theory.

Our solution can thus be interpreted as a rotation-induced dyonic electromagnetized Kerr black hole and the horizon condition is $\Delta = 0$, as in the original Kerr black hole, with the inner and outer horizon determined solely by the parameters $(\mu,a)$.  However, there are  obstacles to the study the corresponding black hole thermodynamics. The asymptotic region of the Melvin-type solutions are not asymptotic Minkowski spacetime and hence we do not have a fiducial frame to fix the time coordinate. We can perform a general coordinate transformation \cite{Ma:2026otg,Ma:2026gdn}
\be
t\rightarrow t'= \lambda_1 t\,,\qquad \varphi \rightarrow \varphi'=\varphi + \lambda_2 t\,,\label{lambda12}
\ee
without changing $g_{\phi\phi}$ and the $\delta\phi=2\pi$ condition. The free parameters $(\lambda_1,\lambda_2)$ make the definition of mass and temperature, etc., ambiguous. Although the first law and Christodoulou-Ruffini relation \cite{Christodoulou:1971pcn} can potentially be used to fix these parameters, as demonstrated in \cite{Ma:2026otg,Ma:2026gdn}, the rotation-induced dyonic charges renders the thermodynamic analysis less informative. Since these charges are not free parameters, they cannot be varied independently of other thermodynamic quantities such as the angular momentum, etc. We therefore shall not attempt to establish the first law here, untill a more general solution with independent charge parameters is constructed.

It is worth pointing out that the induced dyonic charges \eqref{dyoniccharges} are invariant under the transformation \eqref{lambda12}, and hence independent of $(\lambda_1,\lambda_2)$ parameters. It can be established that both the entropy and angular momentum are also independent of $(\lambda_1,\lambda_2)$. The entropy, associated with both inner and outer horizons, are given by
\be
S_\pm=\pi(1-4\mu^2a^2B^2E^2)(r_\pm^2+a^2)\,.
\ee
Computing the angular momentum via the generalized Komar integration \cite{Liu:2022wku} associated with the Killing vector $\xi=\partial/\partial\varphi$ is more involved. We define the dual 1-form field from the equations of motion of the KK theory
\be
d\big(e^{\sqrt{3}\phi}{*F}_{\2}\big)=0\qquad \Rightarrow\qquad e^{\sqrt{3}\phi}{*F}_{\2}=d\mathcal{A}_{\1}.
\ee
The manifestly electromagnetic-dual closed Komar 2-form, satisfying $d{\bf Q}=0$, can be written as
\be
\mathbf{Q}=-{*d}\xi-\frac{1}{2}e^{\sqrt{3}\phi}{*F}_{\2}(i_\xi A_{\1})
+\frac{1}{2}F_{\2}(i_\xi \mathcal{A}_{\1})\,,
\ee
or equivalently,
\be
Q_{\mu\nu}=-2\nabla_{[\mu}\xi_{\nu]}-\frac{1}{2}e^{\sqrt{3}\phi}F_{\mu\nu}
A_{\lambda}\xi^\lambda-\frac{1}{2}{*F}_{\mu\nu}\mathcal{A}_{\lambda}\xi^\lambda\,.
\ee
The corresponding dual one-form of the original one in \eqref{Kerr BE} can be computed, given by
\bea
\mathcal{A}_{\1}&=&-\frac{1}{H_2}\Big[
B\mathbb{L}(1+\frac{BE}{2}\mathbb{L})+\frac{E(1-x^2)\mathbb{P}}{\Sigma}
\big(1+\frac{B^2(1-x^2)\mathbb{P}}{2\Sigma}\big)
\Big]\Big(P_0\,d\varphi+\mathbb{T}dt\Big)\nn\\
&&-B\Big(\mathbb{Z}+\frac{BE}{2}\mathbb{M}\Big)dt\,.
\eea
From this we can compute the angular momentum via the Killing vector \(\partial_\varphi\), and likewise the area of the black hole gives the black hole entropy
\be
J=(1-4\mu^2a^2B^2E^2)\mu a\,.
\ee
Remarkably, the relation of the angular momentum and entropy of the Kerr black hole, namely
\be
S_+ S_-=4\pi^2 J^2\,,\label{spsmJ}
\ee
continue to hold precisely, even after the electromagnetization.

Finally, we note an interesting feature that the rotation prevents the dilaton from decoupling even when $B^2=E^2$. This is because, in the presence of rotation, $F^2$ involves a term proportional to $a E B$, in addition to the $E^2-B^2$ terms. Instead, when $B^2=E^2$, the dilaton becomes odd under reflection across the equatorial plane,
\be
\phi(r,-x)=-\phi(r,x),
\qquad
\phi(r,0)=0\,.
\ee

\section{Electromagnetized Kerr black hole in EMDA theory}
\label{sec:emda}

EMDA theory is commonly regarded as the four-dimensional low-energy effective theory arising from heterotic string theory. It is particularly notable for its strong-weak coupling duality, initially investigated by Sen \cite{Sen:1994fa}. The bosonic sector contains the metric, a Maxwell field, a complex scalar $\tau=\chi + {\rm i} e^{\phi}$. The Lagrangian is
\bea
\mathcal{L}&=&\sqrt{-g} \Big(R-\frac{1}{2}(\partial\phi)^2-\frac{1}{2}e^{-2\phi}(\partial\chi)^2-\frac{1}{4}e^\phi F^2-\frac{1}{4}\chi F_{\mu\nu}\tilde{F}^{\mu\nu}\Big)\,,\quad  \tilde{F}_{\mu\nu}=\frac{1}{2}\epsilon_{\mu\nu\rho\sigma}F^{\rho\sigma}\,.
\eea
Here we adopt the convention$\sqrt{-g}\,\epsilon^{0123}=-1$. Equivalently, the expression above can be rewritten in the language of differential forms as
\be
\mathcal{L}=R\, {*\oneone}-\frac{1}{2}{*d}\phi\wedge d\phi-\frac{1}{2}e^{-2\phi}{*d}\chi\wedge d\chi-\frac{1}{2}e^{\phi}{*F}_{\2}\wedge F_{\2}+\frac{1}{2}\chi F_{\2}\wedge F_{\2}\,.
\ee
The axionic formulation is obtained by Hodge dualizing the three-form field strength $H_{\3}$ in the four-dimensional heterotic string effective action to a pseudoscalar. The hidden symmetry of EMDA theory is $Sp(4,\mathbb{R})$. We first make this symmetry manifest through dimensional reduction and then exploit it to generate solutions immersed in simultaneous external electric and magnetic fields.

\subsection{$Sp(4,\mathbb{R})$ symmetry}

Since our first aim is to introduce external fields rather than charges, we employ the magnetic Ernst transformation in this section. Accordingly, the dimensional reduction is carried out along a spacelike Killing direction. We therefore adopt the following reduction ansatz
\bea
ds_4^2=e^{\Phi}(dz+\bar{\mathcal{A}}_{\1})^2+e^{-\Phi}d\bar{s}_3^2\,,\qquad A_{\1}=\bar{A}_{\1}+X(dz+\bar{\mathcal{A}}_{\1})\,.\label{magnetic Ernst EMDA}
\eea
Introducing the following notation
\be
F_{\2}=dA_{\1}=\bar{F}_{\2}+dX\wedge(dz+\bar{\mathcal{A}}_{\1})\,,\qquad
\bar{F}_{\2}=d\bar{A}_{\1}+X\bar{\mathcal{F}}_{\2}\,,\qquad \bar{\mathcal{F}}_{\2}=d\bar{\mathcal{A}}_{\1}\,,
\ee
we obtain the three-dimensional action
\bea
\bar{\mathcal{L}}&=&\bar{R}\,{\bar{*}\oneone}-\frac{1}{2}\,\bar{*} d\Phi\wedge d\Phi-\frac{1}{2}\,\bar{*}d\phi\wedge d\phi-\frac{1}{2}e^{-2\phi}\,\bar{*}d\chi\wedge d\chi-\frac{1}{2}e^{\phi-\Phi}\,\bar{*}dX\wedge dX\cr
&&-\frac{1}{2}e^{2\Phi}\,\bar{*}\bar{\mathcal{F}}_{\2}\wedge \bar{\mathcal{F}}_{\2}-\frac{1}{2}e^{\phi+\Phi}\,\bar{*}\bar{F}_{\2}\wedge \bar{F}_{\2}+\chi dX\wedge \bar{F}_{\2}\,.\label{3D action EMDA}
\eea
By means of the Bianchi identity, we introduce the following Lagrange-multiplier term
\be
\Delta \bar{\mathcal{L}}=\sigma (d\bar{F}_{\2}-dX\wedge \bar{\mathcal{F}}_{\2})+\psi d\bar{\mathcal{F}}_{\2}\,,
\ee
and recast the theory into a first-order form. The equations of motion for the two two-form fields can then be solved to give
\be
e^{\phi+\Phi}\,\bar{*}\bar{F}_{\2}+d\sigma-\chi dX=0\,,\qquad e^{2\Phi}\,\bar{*}\bar{\mathcal{F}}_{\2}+d\psi+\sigma dX=0\,
\ee
Solving for the two-form fields and substituting them into the first-order formulation eliminates all two-form fields and yields an Einstein-scalar theory
\bea
\bar{\mathcal{L}}&=&\bar{R}\,\bar{*}\oneone-\frac{1}{2}\,\bar{*}d\Phi\wedge d\Phi-\frac{1}{2}\,\bar{*}d\phi\wedge d\phi-\frac{1}{2}e^{-2\phi}\,\bar{*}d\chi\wedge d\chi-\frac{1}{2}e^{\phi-\Phi}\,\bar{*}dX\wedge dX\cr
&&-\frac{1}{2}e^{-\phi-\Phi}\,\bar{*}(d\sigma-\chi dX)\wedge (d\sigma-\chi dX)-\frac{1}{2}e^{-2\Phi}\,\bar{*}(d\psi+\sigma dX)\wedge(d\psi+\sigma dX)\,.\label{3D action scalar EMDA}
\eea
Since the four-dimensional theory already contains an axion, the reduced theory possesses one additional scalar compared with the KK theory in \eqref{3D action scalar}. For later convenience, we introduce two symmetric $2\times2$ matrices, $P$ and $Q$
\be
P=\renewcommand{\arraystretch}{1.15}
\setlength{\arraycolsep}{6pt}
\begin{pmatrix}
e^\Phi+\frac{1}{2}e^\phi X^2 & \frac{1}{\sqrt{2}}e^\phi X \\
\frac{1}{\sqrt{2}}e^\phi X & e^\phi
\end{pmatrix}\,,\qquad
Q=\renewcommand{\arraystretch}{1.15}
\setlength{\arraycolsep}{6pt}
\begin{pmatrix}
\psi+X\sigma-\frac{1}{2}\chi X^2 & \frac{1}{\sqrt{2}}(\sigma-\chi X) \\
\frac{1}{\sqrt{2}}(\sigma-\chi X) & -\chi
\end{pmatrix}.
\ee
We then introduce a \(4\times4\) matrix \(S\),
\be
S=
\setlength{\arraycolsep}{8pt}
\begin{pmatrix}
\mathbb{I}_2 & 0 \\
Q & \mathbb{I}_2
\end{pmatrix}
\begin{pmatrix}
P^{-1} & 0 \\
0 & P
\end{pmatrix}
\begin{pmatrix}
\mathbb{I}_2 & Q \\
0 & \mathbb{I}_2
\end{pmatrix}\,,\label{matrix S EMDA}
\ee
where \(\mathbb{I}_2\) denotes the \(2\times2\) identity matrix. The action \eqref{3D action scalar EMDA} can then be written in the same compact form as \eqref{eincoset}. The scalar sector therefore possesses an \(Sp(4,\mathbb{R})/U(2)\) coset structure. We now summarize the defining condition for the \(Sp(4,\mathbb{R})\) symmetry. For a real \(4\times4\) matrix \(\Gamma\), if it preserves the symplectic form \(\mathbb{J}\),
\be
\Gamma^T \mathbb{J} \Gamma = \mathbb{J},\qquad \mathbb{J}=
\setlength{\arraycolsep}{8pt}
\begin{pmatrix}
0 & \mathbb{I}_2 \\
-\mathbb{I}_2 & 0
\end{pmatrix}\label{Sp4R}
\ee
then \(\Gamma\) is an element of \(Sp(4,\mathbb{R})\). Since the matrix $S$ in \eqref{matrix S EMDA} is symmetric, $S^T=S$, and satisfies $S\mathbb{J}S=\mathbb{J}$, it follows that $S$ itself belongs to $Sp(4,\mathbb{R})$. The transformation matrix $\Gamma(B,E)$ that introduces simultaneous external electric and magnetic fields is
\be
\Gamma(B,E)=
\begin{pmatrix}
1 \hspace{8pt} & \frac{B}{\sqrt{2}} & 0 & -\frac{E}{\sqrt{2}} \\
0 \hspace{8pt} & 1 & -\frac{E}{\sqrt{2}} & 0 \\
0 \hspace{8pt} & 0 & 1 & 0 \\
0 \hspace{8pt} & 0 & -\frac{B}{\sqrt{2}} & 1
\end{pmatrix}.
\ee
It is straightforward to verify that $\Gamma(B,E)$ satisfies the symplectic condition \eqref{Sp4R}. Therefore, under the transformation $S\rightarrow S'=\Gamma S\Gamma^T$, the action \eqref{eincoset} remains invariant.

\subsection{Electromagnetized vacuum and Kerr black hole}

Having established the solution-generating procedure, we apply it to the Minkowski vacuum in cylindrical coordinates \eqref{Mink}. The resulting electromagnetized vacuum is given by
\bea
ds^2&=&\Lambda\Big(-dt^2+d\rho^2+dz^2\Big)+\Lambda^{-1}\rho^2d\varphi^2\,,\cr
A_{\1}&=&-2Ez dt-\frac{B\rho^2}{\Lambda}d\varphi\,,\qquad \phi=\log\frac{(B^2+E^2)\Lambda}{B^2+E^2\Lambda^2}\,,\cr
\chi&=&\frac{BE(1-\Lambda^2)}{B^2+E^2\Lambda^2}\,,\qquad \Lambda=1+\frac{1}{2}(B^2+E^2)\rho^2\,.
\eea
Likewise, if we apply the electromagnetization procedure to the Kerr black hole \eqref{Kerr}, the resulting solution can be written as
\bea
ds^2&=&H_1\Sigma\Big[-\frac{\Delta}{\mathbb{P}}dt^2+\frac{dr^2}{\Delta}+\frac{dx^2}{1-x^2}
\Big]+\frac{(1-x^2)\mathbb{P}}{H_1\Sigma}\Big(d\varphi-\frac{2\mu ar}{\mathbb{P}}dt\Big)^2\,,\cr
A_{\1}&=&\frac{E\mathbb{T}-B\mathbb{L}}{H_1}\Big(d\varphi-\frac{2\mu ar}{\mathbb{P}}dt\Big)+E\mathbb{M}dt\,,\qquad \phi=\log\frac{H_1}{H_2}\,,\cr
\chi&=&\frac{1}{H_2}\Big[
\frac{E^2-B^2}{2}\mathbb{T}-BE\mathbb{L}-\frac{BE(B^2+E^2)}{4}(\mathbb{L}^2+\mathbb{T}^2)
\Big]\,,\label{emdakerr}
\eea
where
\bea
H_1&=&1+\frac{B^2+E^2}{2}\mathbb{L}\,,\qquad H_2=1+BE\mathbb{T}+E^2\mathbb{L}+\frac{E^2(B^2+E^2)}{4}(\mathbb{L}^2+\mathbb{T}^2)\,,\cr
\mathbb{M}&=&-\frac{2x\Delta}{\mathbb{P}}\big[r\Sigma-\mu a^2(1+x^2)\big]\,,\qquad \mathbb{L}=\frac{1-x^2}{\Sigma}\mathbb{P}\,,\cr
\mathbb{P}&=&(r^2+a^2)^2-a^2(1-x^2)\Delta\,,\qquad \mathbb{T}=-2\mu ax\Big(3-x^2+\frac{a^2(1-x^2)^2}{\Sigma}\Big)\,.
\eea
For $E^2=B^2$, we find that the complex scalar $\tau$ satisfies the relation $\tau(r,-x)=1/\overline{\tau}(r,x)$. In the further static limit $a=0$, the axion $\chi$ and the dilaton $\phi$ satisfy the simpler relation $\bar\tau\tau=\chi^2+e^{2\phi}=1$.

We note that the azimuthal angle \(\varphi\) already has the standard period \(2\pi\), \(\varphi\sim\varphi+2\pi\), and therefore requires no rescaling. We can compute the rotation-induced electromagnetic charges of the black hole
\be
Q_e=\frac{1}{16\pi}\int \big(e^{\phi}{*F}_{\2}-\chi F_{\2}\big)=\mu aB\,,
\qquad Q_m=\frac{1}{16\pi}\int F_{\2}=\mu aE\,.
\ee
To determine the angular momentum, we first define the dual 1-form of the EMDA theory
\be
e^\phi {*F}_{\2}-\chi F_{\2}=d\mathcal{A}_{\1}\,,
\ee
and thereby define the generalized closed Komar 2-form:
\be
\mathbf{Q}=-{*d}\xi-\frac{1}{2}(e^\phi {*F}_{\2}-\chi F_{\2})(i_\xi A_{\1})+\frac{1}{2}F_{\2}(i_\xi \mathcal{A}_{\1})
\ee
or equivalently,
\be
Q_{\mu\nu}=-2\nabla_{[\mu}\xi_{\nu]}-\frac{1}{2}(e^{\phi}F_{\mu\nu}+\chi*F_{\mu\nu})
A_{\lambda}\xi^\lambda-\frac{1}{2} {*F}_{\mu\nu}\mathcal{A}_{\lambda}\xi^\lambda\,.
\ee
We find that the dual 1-form given by \eqref{emdakerr} is
\be
\mathcal{A}_{\1}=\frac{B\mathbb{T}+E\mathbb{L}}{H_1}\Big(d\varphi-\frac{2\mu ar}{\mathbb{P}}dt\Big)+B\mathbb{M}dt\,.
\ee
Using this result, we obtain the angular momentum associated with the Killing vector $\xi=\partial/\partial\varphi$. The entropy associated with both inner and outer horizons can also be easily computed. We find
\be
J=\mu a\,,\qquad S_\pm=\pi(r_\pm^2+a^2)\,.
\ee
These are identical to the original Kerr black hole. Note that all the quantities we derived, $Q_e$, $Q_m$, $S_\pm$ and $J$ are independent of the transformation \eqref{lambda12}.

\section{Conclusions}
\label{sec:conclusions}

In this paper, we constructed rotating black hole solutions immersed in both electric and magnetic external fields, starting from the Ricci-flat Kerr black hole and exploiting the hidden $SL(3,\mathbb{R})$ and $Sp(4,\mathbb{R})$ symmetries of KK and EMDA theories, respectively. We also investigated their basic properties, calculated their angular momenta and horizon entropies and determined the rotation-induced electric and magnetic charges in both theories. We therefore refer to these solutions as dyonic Melvin-type electromagnetized Kerr black holes with rotation-induced charges.
The emergence of the induced electric and magnetic charges is analogous to that in the Kerr-Melvin black hole of Einstein-Maxwell theory. A remarkable result is that the entropy/angular momentum relation \eqref{spsmJ} remains unchanged under the electromagnetization.

Since the KK theory arises from the circle reduction of pure gravity in five dimension, this five-to-four route provides a powerful solution-generating technique for constructing electromagnetized black holes. However, a naive application of this procedure will generate a twist term known as Taub-swirling. To understand the origin of this, we explored the alternative four-to-three route, which is based on the fact that the KK theory exhibits an $SL(3,\mathbb{R})$ hidden symmetry upon dimensional reduction to three dimensions. We found that generating a vacuum with both external electric and magnetic fields was considerably more subtle than generating one with either field alone. Using Minkowski spacetime in cylindrical coordinates as a simple example, we demonstrated that a generic $SL(3,\mathbb{R})$ transformation can introduce the nontrivial twist term. Under spacelike reduction, which generates external electromagnetic fields, this gives rise to a Taub-swirling structure, whereas under timelike reduction, which generates electric and magnetic charges, it leads to a Taub-NUT structure. Rasheed avoided this difficulty in constructing dyonic black holes by choosing a special $SL(3,\mathbb{R})$ group element. Similarly, we identified an appropriate group element that preserves the diagonal form of the electromagnetic vacuum. We then reformulated this specific $SL(3,\mathbb R)$ transformation as a modified five-to-four procedure, which generates the same KK solutions without either Taub-swirling or Taub-NUT twist terms.

Our modified solution-generating technique allows one to construct solutions with independent dyonic charges, as well as independent external electromagnetic fields in the KK theory or EMDA theories. In this paper, however, we are primarily concerned with electromagnetization alone, and the resulting electromagnetized Kerr black hole carries induced electric and magnetic charges by rotation. These solutions therefore do not describe genuinely neutral rotating black holes immersed in external electromagnetic fields in the sense of \cite{Ovcharenko:2026tos}. It is of great interest to investigate whether a genuine neutral rotating black holes requires new Ricci-flat seeds such as those constructed in \cite{Ma:2026otg,Ma:2026gdn}, or whether new solutions-generating techniques are required.

\section*{Acknowledgement}

L.M.~is supported in part by National Natural Science Foundation of China (NSFC) grant No.~12447138, Postdoctoral Fellowship Program of CPSF Grant No.~GZC20241211, the China Postdoctoral Science Foundation under Grant No.~2024M762338, the National Key Research and Development Program No.~2022YFE0134300, and also by the Tianjin University Self-Innovation Fund New Inspiration Fund No.~2026XQM09934. H.L.~is supported in part by the NSFC grants No.~12375052 and No.~11935009. The work is also supported in part by the Tianjin University Self-Innovation Fund Extreme Basic Research Project Grant No.~2025XJ21-0007. We also made an extensive use of AI programs in both research and English proofreading.

\appendix
\section{Dyonic KK black hole from the five-to-four route}
\label{sec:kkdyon}

In the main text, we discussed the electromagnetization of Minkowski vacuum in the KK theory, and showed that the naive application can create an electromagnetic field dependent ($EB$) component of swirling. It is well known that the charged KK black hole can be generated by applying a Lorentz boost along the time of the Kerr black hole and the $y$ of the fifth dimension. Similarly, naive application of the five-to-four route of solution generating technique will give rise to the dyonic black hole with a Taub-NUT component that is proportional to $q_e q_m$, the product of electric and magnetic charge parameters. Such Taub-NUT component was avoided by the four-to-three route in \cite{Rasheed:1995zv} for the more general case of rotating black holes. (See also \cite{Larsen:1999pp}.)

Following the strategy used for electromagnetization in the main text, we now generate the dyonic black hole without Taub-NUT via the modified five-to-four route. To illustrate this procedure, we take the Schwarzschild black hole as the Ricci-flat seed:
\be
ds^2_4=-fdt^2+f^{-1}dr^2+r^2d\Omega_2^2\,,\qquad f=1-\frac{2\mu_0}{r}\,.\label{Sch}
\ee
We define the boost-charging procedure as follows: First, we uplift a four-dimensional solution to five dimensions using \eqref{KK reduction 5 to 4}. We then perform a Lorentz boost of the form
\be
t\quad \rightarrow\quad t'=c_it+s_iy\,,\qquad y\quad \rightarrow\quad y'=s_it+c_iy\,.
\ee
Here, $c_i=\cosh\alpha_i$ and $s_i=\sinh\alpha_i$. We then reduce back to four dimensions using the ansatz \eqref{KK reduction 5 to 4}, obtaining an electrically-charged black hole. In analogy with the procedure for introducing an external magnetic field, we denote this sequence of uplifting, performing a Lorentz boost, and dimensional reduction as the boost-charging procedure, denoted by $\mathcal{Q}_i$.

Taking the Schwarzschild black hole \eqref{Sch} as a seed solution, we first apply the boost-charging transformation $\mathcal{Q}_1$ to generate the electrically-charged solution, given by
\bea
ds^2_{4,\1}&=&-\frac{f}{\sqrt{\Lambda}}dt^2+\sqrt{\Lambda}\Big(f^{-1}dr^2+r^2d\Omega_2^2\Big)\,,\qquad A_{\1}^{\1}=\frac{2\mu_0 s_1c_1}{r\Lambda}dt\,,\cr
\phi^{\1}&=&\frac{\sqrt{3}}{2}\log\Lambda\,,\qquad \Lambda=1+\frac{2\mu_0 s_1^2}{r}\,.
\eea
We then apply the electromagnetic duality transformation $\mathcal{D}$ \eqref{ele mag dual}, which leaves the metric unchanged while transforming the Maxwell field and the dilaton as follows
\be
ds^2_{4,\2}=ds^2_{4,\1}\,,\qquad A_{\1}^{\2}=-2\mu_0 s_1c_1xd\varphi\,,\qquad \phi^{\2}=-\frac{\sqrt{3}}{2}\log\Lambda\,.
\ee
At this stage, the electric charge is converted into a magnetic charge. We then proceed with the second boost-charging transformation $\mathcal{Q}_2$, yielding
\bea
ds^2_{4,\3}&=&-\frac{f}{\sqrt{\Lambda\Xi}}(dt+2\mu_0 s_1c_1s_2xd\varphi)^2+\sqrt{\Lambda\Xi}\Big(f^{-1}dr^2+r^2d\Omega_2^2\Big)\,,\cr
A_{\1}^{\3}&=&\frac{s_2c_2(1-f\Lambda)}{\Xi}(dt+2\mu_0 s_1c_1s_2xd\varphi)-2\mu_0 s_1c_1c_2xd\varphi\,,\cr
\phi^{\3}&=&\frac{\sqrt{3}}{2}\log\frac{\Xi}{\Lambda}\,,\qquad \Xi=c_2^2-f\Lambda s_2^2
\eea
We observe that the introduction of both electric and magnetic charges also gives rise to a nontrivial cross term in the metric, corresponding to a NUT-type structure.

In order to remove this Taub-NUT twist, we apply the inverse electromagnetic duality transformation $\mathcal{D}^{-1}$, under which the Maxwell field and the dilaton transform as follows
\bea
ds^2_{4,\4}&=&ds^2_{4,\3}\,,\qquad A_{\1}^{\4}=\frac{2\mu_0 s_1c_1c_2}{r\Lambda}(dt+2\mu_0 s_1c_1s_2xd\varphi)+2\mu_0 s_2c_2(1-s_1^2)xd\varphi\,,\cr
\phi^{\4}&=&\frac{\sqrt{3}}{2}\log\frac{\Lambda}{\Xi}\,.
\eea
Finally, we perform the third boost-charging transformation $\mathcal{Q}_3$, choosing its boost parameters $s_3$ and $c_3$ to eliminate the NUT structure as follows
\be
s_1c_1c_3=c_2(1-s_1^2)s_3\,.
\ee
Next, we introduce a new set of parameters
\bea
\nu&=&\sqrt{1-s_e^2s_m^2}\,,\qquad \mu_0=\mu\nu\,,\qquad \mathcal{K}=\sqrt{\nu^2+c_e^2(1+2\nu)}\,,\cr
c_1&=&\frac{c_e\sqrt{2(1+\nu)}}{\mathcal{K}}\,,\qquad s_1=-\frac{s_ec_m}{\mathcal{K}}\,,\qquad c_2=\frac{c_m}{\nu}\sqrt{\frac{1+\nu}{2}}\,,\qquad s_2=\frac{s_m\mathcal{K}}{\nu\sqrt{2(1+\nu)}}\,,\cr
c_3&=&\frac{c_e^2+\nu}{\mathcal{K}}\,,\qquad s_3=-\frac{s_ec_e}{\mathcal{K}}\,,
\eea
to rewrite the final result in a more compact form
\bea
ds_4^2&=&-\frac{f}{\sqrt{H_1H_2}}dt^2+\sqrt{H_1H_2}\Big(f^{-1}dr^2+r^2d\Omega_2^2\Big)\,,\cr
A_{\1}&=&-\frac{2\mu s_ec_e}{rH_2}\Big[1+\frac{\mu(c_m^2-\nu)}{r}\Big]dt+2\mu s_mc_mxd\varphi\,,\qquad \phi=\frac{\sqrt{3}}{2}\log\frac{H_2}{H_1}\,,\cr
H_1&=&1+\frac{2\mu(c_m^2-\nu)}{r}+\frac{2\mu^2c_m^2(1-\nu)}{r^2}\,,\qquad
H_2=1+\frac{2\mu(c_e^2-\nu)}{r}+\frac{2\mu^2c_e^2(1-\nu)}{r^2}\,,\cr
f&=&1-\frac{2\mu\nu}{r}\,.
\eea
This solution is the Gibbons-Wiltshire black hole, whose complete thermodynamic properties are given by
\bea
M&=&\frac{\mu(c_e^2+c_m^2)}{2}\,,\qquad T=\frac{\nu}{4\pi\mu c_ec_m(1+\nu)}\,,\qquad S=2\pi\mu^2c_ec_m(1+\nu)\,,\cr
Q_e&=&\frac{\mu s_ec_e}{2}\,,\qquad Q_m=\frac{\mu s_mc_m}{2}\,,\qquad \Phi_e=\frac{s_e(c_m^2+\nu)}{c_e(1+\nu)}\,,\qquad
\Phi_m=\frac{s_m(c_e^2+\nu)}{c_m(1+\nu)}\,.
\eea
We therefore demonstrate that we can derive the Gibbons-Wiltshire black hole without going down to three dimensions.

\section{Taub-NUT and Taub-swirling structure from $SL(3,\mathbb R)$}
\label{sec:nutswirling}

In Section \ref{sec:emvacuum}, we saw that an \(SL(3,\mathbb{R})\) transformation group element \(\Gamma''(B,E)\) \eqref{Gamma dd} is required to eliminate the swirling structure. It is therefore worthwhile to examine the properties of this transformation itself. Taking again Minkowski spacetime in four-dimensional cylindrical coordinates \eqref{Mink} as the seed solution, we consider a general transformation \(\Gamma''(c)\)
\be
\Gamma''(c)=\renewcommand{\arraystretch}{1.15}
\setlength{\arraycolsep}{7pt}
\begin{pmatrix}
1 & 0 & c \\
0 & 1 & 0 \\
0 & 0 & 1
\end{pmatrix}\label{Gamma c}
\ee
and find that the resulting solution is Ricci-flat, given by
\be
ds_4^2=(1+c^2\rho^4)(-dt^2+d\rho^2+dz^2)+\frac{\rho^2}{1+c^2\rho^4}(d\varphi-4czdt)^2\,,\qquad A_{\1}=0\,,\quad \phi=0\,.
\ee
This is the Taub-swirling solution \cite{Taub:1950ez}; it can likewise be obtained through the \(SU(2,1)\) transformation of ordinary Einstein-Maxwell theory discussed in \cite{Gibbons:2013yq}.

We can also generate the NUT parameter by means of \eqref{Gamma c}. It is important to notice that generating the NUT parameter requires reduction along the timelike Killing direction \(dt\) rather than along the spacelike Killing direction \(d\varphi\) in \eqref{magnetic Ernst} \cite{Gibbons:1987ps}. Moreover, using the Minkowski vacuum \eqref{Mink} as the seed still yields the Minkowski vacuum. Therefore, here we choose the Schwarzschild black hole \eqref{Sch} as the seed, and then perform a coordinate transformation and parameter redefinition
\be
t_N=\frac{t}{\sqrt{1+c^2}}\,,\qquad r_N=\sqrt{1+c^2}r-\frac{2\mu_0 c^2}{\sqrt{1+c^2}}\,,\qquad \mu=\frac{\mu_0(1-c^2)}{\sqrt{1+c^2}}\,,\qquad n=\frac{2\mu_0c}{\sqrt{1+c^2}}\,,
\ee
the resulting solution is precisely the standard Ricci-flat Taub-NUT solution
\be
ds_4^2=-f_N(dt_N+2nxd\varphi)^2+\frac{dr_N^2}{f_N}+(r_N^2+n^2)d\Omega_2^2\,,\quad f_N=\frac{r_N^2-2\mu r_N-n^2}{r_N^2+n^2}\,.
\ee
The standard Taub-NUT solution can also be generated from the Schwarzschild black hole by exploiting
the hidden \(SU(2,1)\) symmetry of Einstein-Maxwell theory in \cite{Gibbons:2013yq}, where the reduction along the \(dt\) direction must again be taken into account, and the corresponding group element is the same as that used to generate the Taub-swirling solution.

\end{document}